\documentclass{article}
\usepackage{spconf,amsmath,graphicx}
\usepackage[colorlinks=true, linkcolor=black, citecolor=black, urlcolor=black]{hyperref}
\usepackage{algorithm}
\usepackage[noend]{algpseudocode}
\usepackage{siunitx}
\usepackage{url}
\usepackage{pifont}
\usepackage{xcolor}
\usepackage{comment}
\usepackage{multirow, makecell}
\usepackage{enumitem}
\usepackage{booktabs}
\usepackage{needspace}

\title{From Independence to Interaction: Speaker-Aware Simulation of Multi-Speaker Conversational Timing}

\name{Máté Gedeon$^{*,\dagger}$ \thanks{© 2025 IEEE. Personal use of this material is permitted. Permission from IEEE must be obtained for all other uses, in any current or future media, including reprinting/republishing this material for advertising or promotional purposes, creating new collective works, for resale or redistribution to servers or lists, or reuse of any copyrighted component of this work in other works.}, Péter Mihajlik$^{*}$}
\address{
    $^{*}$Dept. of Telecommunications and Artificial Intelligence, \\
    Budapest University of Technology and Economics, Hungary \\
    $^{\dagger}$Speechtex Ltd. \\
    \texttt{gedeonm@edu.bme.hu, mihajlik@tmit.bme.hu}
}

\begin{document}
%
\maketitle

\begin{abstract}
We present a speaker-aware approach for simulating multi-speaker conversations that captures temporal consistency and realistic turn-taking dynamics. Prior work typically models aggregate conversational statistics under an independence assumption across speakers and turns. In contrast, our method uses speaker-specific deviation distributions enforcing intra-speaker temporal consistency, while a Markov chain governs turn-taking and a fixed room impulse response preserves spatial realism. We also unify pauses and overlaps into a single gap distribution, modeled with kernel density estimation for smooth continuity. Evaluation on Switchboard using intrinsic metrics—global gap statistics, correlations between consecutive gaps, copula-based higher-order dependencies, turn-taking entropy, and gap survival functions—shows that speaker-aware simulation better aligns with real conversational patterns than the baseline method, capturing fine-grained temporal dependencies and realistic speaker alternation, while revealing open challenges in modeling long-range conversational structure.
\end{abstract}

\section{Introduction}
\label{sec:intro}
Processing multi-speaker conversational speech is crucial for applications such as meeting transcription and voice assistants, where both accurate transcription and diarization (\textit{who spoke when}) are required \cite{diarization_review}. End-to-end architectures achieve strong performance in these tasks but rely on large volumes of annotated conversational data \cite{Chen2020}, which remain scarce—particularly for low-resource languages and specialized domains.

ASR systems trained on single-speaker data often perform well under ideal conditions but degrade significantly in overlapping speech, which is typically absent from training corpora \cite{Yu2016PIT}. A common solution to this scarcity is the generation of synthetic conversations from single-speaker corpora \cite{Bartelds2023}. Early methods created \textit{simulated mixtures} by concatenating utterances with random pauses \cite{Fujita2019}, which are computationally simple but produce unnatural turn-taking and overlap patterns. Later approaches improved realism by sampling pause and overlap statistics from real conversations \cite{Landini2022,Yamashita2022Naturalness}, significantly boosting diarization performance \cite{Landini2022MultiSpeakerEEND}, but they still treat speakers independently, rely on general distributions, and fail to capture complex conversational dynamics. Synthetic conversation generation has also improved ASR robustness by creating realistic overlapping scenarios from single-speaker corpora \cite{LibriheavyMix,SOT, Yang2023}, which is particularly valuable for low-resource languages and specialized domains where natural multi-talker data is prohibitively expensive or infeasible to collect. However, such synthetic data may still fail to capture the full range of spontaneous conversational dynamics, limiting its effectiveness in truly natural interaction settings.

Recent end-to-end neural diarization (EEND) approaches \cite{Fujita2019, Horiguchi2020, Kinoshita2021Advances} illustrate both the potential and the challenge: by replacing complex pipelines with a single neural model, they show that rich conversational patterns can be learned directly. However, like ASR, their effectiveness is constrained by the lack of realistic multi-speaker training data. This further underscores the importance of advancing synthetic conversation generation—not just as a data augmentation technique, but as a prerequisite for progress in end-to-end conversational speech processing.


Synthetic data generation has therefore become indispensable for both diarization and conversational ASR. Random concatenation \cite{Fujita2019} is a simple method but yields artificial dialogue dynamics. Statistical modeling \cite{Landini2022, Zulkashev2023, Park2023} represented a paradigm shift, as it extracts pause, overlap, and transition statistics from real conversations to better capture collaborative turn-taking. However, these models still rely on general histograms, causing independence between a specific speaker's utterances and introduce discretization artifacts. More sophisticated density estimation and conditional modeling are needed to bridge this gap.

The key contributions of this work are as follows:
\begin{itemize}
    \item We propose a speaker-aware conversation simulation framework that preserves speaker-specific timing traits throughout a conversation.
    \item We introduce the use of kernel density estimation (KDE) \cite{KDE} for non-parametric modeling of conversational gaps and overlaps, combining the advantages of data-driven histograms and parametric models.
    \item We incorporate a Markov-based turn-taking transition model and room-consistent acoustic simulation to further improve conversational realism.
    \item We demonstrate that the framework works with a non-conversational single-speaker dataset, enabling broader applicability for data augmentation.
    \item We propose intrinsic evaluation metrics for similarity between simulated and real conversations, providing standardized tools to support future research in synthetic conversation modeling.
\end{itemize}
Upon publication, we will release the code used to generate the evaluation dataset, along with the corresponding KDE models.

\section{Methodology}
\subsection{Simulated conversations}
Landini et al.~\cite{Landini2022} proposed simulated conversations to address a key limitation of traditional mixtures: the independent treatment of speakers, neglecting the collaborative nature of dialogue. Their method derives statistics from real conversations, but relies on \emph{general} rather than speaker-specific distributions, leaving residual independence assumptions and limited conversational realism.

The approach is based on four statistics: (1) pause length distributions within a speaker ($D_{=\text{speaker}}$), (2) pause length distributions across speakers ($D_{\neq\text{speaker}}$), (3) overlap length distributions ($D_{\text{overlap}}$), and (4) the probability $p=\frac{ds}{ds+ov}$ of pause vs. overlap in cross-speaker transitions, where $ds$ and $ov$ denote counts of pauses and overlaps. While these variables enrich dialogue modeling beyond concatenation, they remain fragmented across multiple distributions.

Utterances are sampled without replacement so each original conversation is used once, with segments randomly interleaved while preserving each speaker's order. This ensures speaker coherence but not per-speaker adaptation of timing. Gap (silence) insertion depends on transition type: same-speaker transitions sample from $D_{=\text{speaker}}$, cross-speaker transitions from $D_{\neq\text{speaker}}$ or $D_{\text{overlap}}$, chosen with probability $p$. Thus, overall statistics are captured, but speaker-specific timing traits are not.

Compared to mixtures, this method better matches real conversations in silence percentages, single-speaker ratios, and overlap distributions, yet still lacks within-speaker temporal consistency. Standard augmentations (noise, reverberation) are also applied, though the latter does not reflect conversational acoustics realistically.

\subsection{Speaker-aware simulated conversations}
The original simulated conversation framework improves realism compared to mixtures but suffers from overly simplistic statistical assumptions. In particular, it treats all speakers identically by sampling from \emph{general} distributions, which ignores the fact that conversational behavior tends to be temporally consistent within each speaker. For example, a participant who frequently leaves short gaps early in a dialogue is likely to maintain similar timing patterns later on.

To address this, first we unify conversational dynamics into a single distribution where negative gap values correspond to overlaps, non-negative gaps correspond to pauses at different-speaker transitions, and the integral over the negative domain equals $p_{\text{overlap}}$. This compact representation eliminates redundant variables and simplifies the algorithm, though it still cannot fully capture higher-order conversational dependencies.

Previous work has assumed exponential distributions with approximated parameters \cite{Yamashita2022Naturalness} or relied on histogram-based sampling \cite{Landini2022}. While these represent parametric and non-parametric strategies, respectively, both approaches have limitations: parametric models impose restrictive functional forms, and histograms suffer from discretization artifacts. In our work, we adopt a non-parametric approach but replace histogram-based modeling with \emph{Kernel Density Estimation} (KDE) \cite{KDE}. KDE yields smooth, continuous density functions that better capture conversational timing patterns, avoids discretization issues, and incorporates estimation uncertainty. Since empirical gap distributions are skewed, we applied the Yeo–Johnson transformation \cite{Yeo-Johnson} to make the data more Gaussian-like, which improves KDE accuracy using a Gaussian kernel.

Beyond adopting KDE, we introduce \emph{speaker-awareness} through a fixed \emph{speaker deviation distribution}. For each speaker, the initial gap is sampled from a distribution containing mean gaps of speakers (estimated from training data), while subsequent values are obtained by adding deviations drawn from the speaker deviation distribution. This design preserves realism in initial turns and enforces temporal consistency across subsequent turns.

We further incorporate a \emph{turn-taking transition matrix}, implemented as a Markov chain \cite{Yamashita2022Naturalness}, which governs speaker changes in a statistically grounded manner and increases interaction naturalness. Depending on the application scenario and the amount of available data, the Markov chain can be extended to an $n$-th order formulation, providing flexibility.

Finally, while the baseline approach assigns independent room impulse responses (RIRs) to each speaker, causing unnatural spatial inconsistency, we resolve this by fixing a single room per simulated conversation and assigning distinct positions to each speaker. This ensures spatial realism, though at the cost of reduced acoustic diversity across generated samples. Algorithm \ref{alg:conversation_kde} formalizes our approach \footnote{$MixAudio(G, y_n, \delta_n)$ concatenates $y_n$ to $G$, with gap $\delta_n$}.  

\begin{algorithm}[h]
\caption{Speaker-aware conversation simulation}
\label{alg:conversation_kde}
\hspace*{\algorithmicindent} \textbf{Input:} 
$\mathcal{S}$ \Comment{\footnotesize Set of available speakers} \\
\hspace*{\algorithmicindent} \hspace*{\algorithmicindent} 
$\mathcal{U} = \{ U_s \}_{s \in \mathcal{S}}$ \Comment{\footnotesize Utterances per speaker $s$} \\
\hspace*{\algorithmicindent} \hspace*{\algorithmicindent} 
$\mathcal{N}, \mathcal{R}$ \Comment{\footnotesize Background noise signals, Possible SNR values} \\
\hspace*{\algorithmicindent} \hspace*{\algorithmicindent} 
$\mathcal{I}$ \Comment{\footnotesize Room impulse responses (RIRs)} \\
\hspace*{\algorithmicindent} \hspace*{\algorithmicindent} 
$N_{\mathrm{spk}}, N_{\mathrm{u}}$ \Comment{\footnotesize Number of speakers/utterances per conversation} \\
\hspace*{\algorithmicindent} \hspace*{\algorithmicindent} 
$P_{\mathrm{turn}}$ \Comment{\footnotesize Markov transition matrix for turn-taking} \\
\hspace*{\algorithmicindent} \hspace*{\algorithmicindent} 
$\hat{D}_{=}$, $\hat{D}_{\neq}$ \Comment{\footnotesize Mean pause distributions: same/different speaker} \\
\hspace*{\algorithmicindent} \hspace*{\algorithmicindent} 
$V_{=}$, $V_{\neq}$ \Comment{\footnotesize Zero-mean speaker deviation distributions}

\begin{algorithmic}[1]
\State $G \gets \emptyset$ \Comment{\footnotesize Processed audio segments}
\State $\mathcal{S}' \gets \mathrm{SampleSubset}(\mathcal{S}, N_{\mathrm{spk}})$
\State Assign RIR $h_s \in \mathcal{I}$ for each $s \in \mathcal{S}'$ (same room, distinct positions)
\State Initialize empty dictionaries $\mu_s^{same}$, $\mu_s^{diff}$ for base timing values
\State Choose initial speaker $X_1 \sim \mathrm{Uniform}(\mathcal{S}')$
\For{$n \gets 1$ to $N_{\mathrm{u}}$}
    \If{$n > 1$}
        \State $X_n \sim P_{\mathrm{turn}}(X_{n-1}, \cdot)$ \Comment{\footnotesize Sample next speaker}
    \EndIf
    \State $u_n \gets \mathrm{SampleUtterance}(U_{X_n})$
    \State $y_n \gets u_n * h_{X_n}$ \Comment{\footnotesize Apply convolution with fixed RIR}
    \If{$n = 1$}
        \State $G \gets \mathrm{MixAudio}(G, y_n, 0)$
    \ElsIf{$X_n = X_{n-1}$} \Comment{\footnotesize Same speaker}
            \If{$X_n \notin \mu_s^{same}$}
                \State $\mu_s^{same}[X_n] \gets \mathrm{Sample}(\hat{D}_{=})$
                \State $\delta_n \gets \mu_s^{same}[X_n]$
            \Else
                \State $\delta_n \gets \mu_s^{same}[X_n] + \mathrm{Sample}(V_{=})$
            \EndIf
        \ElsIf{$X_n \notin \mu_s^{diff}$} \Comment{\footnotesize Different speaker}
                \State $\mu_s^{diff}[X_n] \gets \mathrm{Sample}(\hat{D}_{\neq})$
                \State $\delta_n \gets \mu_s^{diff}[X_n]$
            \Else
                \State $\delta_n \gets \mu_s^{diff}[X_n] + \mathrm{Sample}(V_{\neq})$
            \EndIf
        \State $G \gets \mathrm{MixAudio}(G, y_n, \delta_n)$
\EndFor
\State Mix $G$ into mono signal $z(t)$
\State Add sampled background noise $n_b \sim \mathcal{N}$ and scale to SNR $r \sim \mathcal{R}$
\end{algorithmic}
\end{algorithm}

\section{Experiments}

Evaluating simulated dialogues is challenging: extrinsic metrics (e.g., ASR or EEND performance) gauge downstream utility, useful for applications, but our aim is to demonstrate value at a more principled, theoretical level. Intrinsic metrics assess similarity to natural conversations but lack standardization. We thus report complementary intrinsic measures capturing both (i) distributional properties and (ii) fine-grained conversational dynamics, focusing on relative fidelity to real data rather than absolute, corpus-dependent scores.

\subsection{Experimental setup}
Conversational statistics were extracted from \textit{Switchboard-1 Release 2 (SB)} \cite{switchboard}, which serves as the primary \emph{target corpus}. To contextualize metric variability, we also include CallHome (CH)~\cite{CallHome}, a contrastive corpus with a similar conversational format but notably different temporal dynamics. CH is not used for model training or generation, but rather to illustrate how evaluation metrics can vary across structurally comparable datasets.

Using statistics derived from SB, we construct two simulated corpora based on LibriTTS~\cite{LibriTTS} speech material:
\begin{itemize}
    \item \emph{Simulated Conversation (SC)}, generated using the baseline simulation method of Landini et al.~\cite{Landini2022}.
    \item \emph{Speaker-aware SC (SASC)}, produced with our proposed speaker-aware variant under identical conditions.
\end{itemize}
Evaluation compares real corpora (SB, CH) with simulated corpora (SC and SASC), both designed to emulate SB. Importantly, the evaluation metrics are distinct from the descriptive features used during generation, allowing for a more independent assessment of simulation quality.

The primary goal of the evaluation phase is to demonstrate improvements in speaker realism and temporal consistency achieved by our speaker-aware model, as reflected in relative metric performance.

\subsection{Global gap statistics}
We first examine inter-turn gaps, i.e., the silence between consecutive segments (negative values indicate overlap). Table~\ref{tab:gapstats} reports mean, median, and standard deviation. Although absolute values differ by corpus, our speaker-aware model (SASC) aligns more closely with SB than SC, indicating improved temporal realism. While SC relies on histograms extracted from SB—so its mean should theoretically be similar—the discrepancy arises from its lack of turn-taking modeling, which causes unrealistic sampling frequencies of same-speaker versus speaker-change transitions. We note that the half-second overlap as mean reflects the underlying annotation, which may not be entirely precise, since such consistent overlapping is uncommon in natural conversations. However, as both methods relied on the same data, the comparison remains fair—though this highlights the need for caution when interpreting statistics derived from the SB annotation.

\begin{table}[h]
\centering
\resizebox{0.9\columnwidth}{!}{
\begin{tabular}{lccc}
\toprule
\textbf{Corpus} & \textbf{Mean Gap (s)} & \textbf{Median Gap (s)} & \textbf{Std. Dev. (s)} \\
\midrule
CH & -0.004 &  0.120 & 1.545 \\
SB (target) & -0.517 & -0.404 & 0.920 \\
\midrule
SC~\cite{Landini2022}     & -0.097 &  0.000 & 0.799 \\
SASC      & -0.619 & -0.680 & 0.835 \\
\bottomrule
\end{tabular}}
\caption{Descriptive statistics of inter-turn gaps.}
\label{tab:gapstats}
\end{table}

\subsection{Local temporal dependencies}
Conversational flow also exhibits local correlations between successive gaps (the duration of one gap is correlated with the next). To quantify this, we compute Pearson’s $r$ \cite{PearsonCorr}, Spearman’s $\rho$ \cite{Spearman1904}, Kendall’s $\tau$ \cite{Kendall}, distance correlation (DCorr) \cite{Szekely2007}, and mutual information (MI) \cite{Shannon} for both speakers, and report the mean values in Table~\ref{tab:gapcorrelations}. For mutual information, we use the uniformized distributions to remove the influence of marginal effects. SASC recovers significantly stronger temporal structure. Interestingly, our method surpasses SB in mutual information, a phenomenon that warrants further investigation.

\begin{table}[h]
\centering
\resizebox{1.0\columnwidth}{!}{
\begin{tabular}{lccccc}
\toprule
\textbf{Corpus} & \textbf{Pearson $r$} & \textbf{Spearman $\rho$} & \textbf{Kendall $\tau$} & \textbf{DCorr} & \textbf{MI} \\
\midrule
CH   & 0.154 & 0.313 & 0.218 & 0.298 & 0.065 \\
SB (target) & 0.074 & 0.101 & 0.097 & 0.100 & 0.003 \\
\midrule
SC~\cite{Landini2022}   & 0.000 & 0.003 & 0.002 & 0.022 & 0.004 \\
SASC        & 0.051 & 0.055 & 0.038 & 0.055 & 0.024 \\
\bottomrule
\end{tabular}}
\caption{Correlation measures between consecutive gaps.}
\label{tab:gapcorrelations}
\end{table}

\subsection{Copula models}
To capture higher-order dependencies, we fit Clayton copulas \cite{Clayton1978} emphasizing dependencies between short gaps and Gumbel copulas focusing on long gaps. Table~\ref{tab:copulas} reports mean log-likelihoods (higher is better). Our method improves over SC across both, though real data remains stronger. This indicates progress in modeling higher-order temporal dependencies, while also highlighting open challenges in fully replicating long-range conversational structure.

\begin{table}[h]
\centering
\resizebox{0.6\columnwidth}{!}{
\begin{tabular}{llc}
\toprule
\textbf{Copula} & \textbf{Corpus} & \textbf{Log-Likelihood} \\
\midrule
\multirow{4}{*}{Clayton} 
  & CH  & 5.042e-02 \\
  & SB (target) & 2.356e-02 \\
  \cmidrule{2-3}
  & SC     & -6.000e-05 \\
  & SASC & 1.194e-03 \\
\midrule
\multirow{4}{*}{Gumbel} 
  & CH   & 4.609e-02 \\
  & SB (target) & 1.708e-02 \\
  \cmidrule{2-3}
  & SC     & -1.162e-02 \\
  & SASC & 9.580e-04 \\
\bottomrule
\end{tabular}}
\caption{Copula mean log-likelihoods for successive gaps.}
\label{tab:copulas}
\end{table}

\subsection{Turn-taking evaluation}
Speaker alternation was measured via average row-wise transition entropy (Table~\ref{tab:turntaking}). SC shows no structure (entropy = 1.0), whereas our speaker-aware model closely matches SB, demonstrating the importance of explicit speaker turn modeling.

\begin{table}[h]
\centering
\resizebox{0.55\columnwidth}{!}{
\begin{tabular}{lc}
\toprule
\textbf{Corpus} & \textbf{Turn-Taking Entropy} \\
\midrule
CH & 0.863 \\
SB (target) & 0.950 \\
\midrule
SC     & 1.000 \\
SASC       & 0.946 \\
\bottomrule
\end{tabular}}
\caption{Average turn-taking entropy; higher indicates more balanced alternation.}
\label{tab:turntaking}
\end{table}

\subsection{Survival curves}
\begin{figure}[h]
    \centering
    \includegraphics[width=0.8\linewidth]{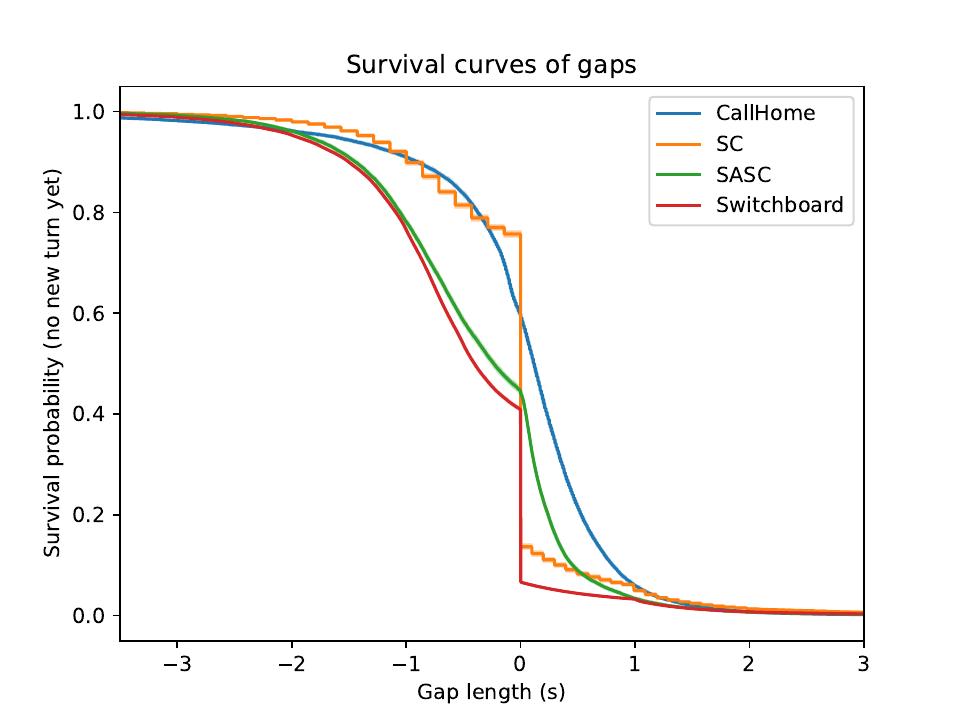}
    \caption{Survival functions $S(t)$ of conversational gaps.}
    \label{fig:survival}
\end{figure}

Gap survival functions $S(t)$, modeling the probability that silence persists longer than $t$ seconds \cite{Kaplan-Meyer} provide full pause distributions (Fig.~\ref{fig:survival}). Our method better approximates SB across both short and long silences compared to SC.

\section{Conclusion}
We presented a speaker-aware extension of simulated conversation generation that unifies conversational gap and overlap modeling, incorporates temporal consistency through speaker deviation distributions, and improves turn-taking realism in terms of the investigated metrics with a Markov-chain framework. Unlike previous approaches relying on fragmented statistics or parametric assumptions, our KDE-based non-parametric modeling yields smooth gap distributions while maintaining statistical flexibility. 

Intrinsic evaluation across gap statistics, local temporal dependencies, copula-based higher-order structures, and survival curves demonstrates that our speaker-aware approach more closely reproduces the statistical properties of natural conversations than the baseline of Landini et al.~\cite{Landini2022}, including improved inter-turn gap correlations, turn-taking entropy, and acoustic consistency. While some challenges remain in modeling long-range conversational structure and low-resource scenarios, our framework provides a robust foundation for downstream speech and dialogue research. 

Moreover, this work opens avenues for future research, such as more robust speaker-specific dependency modeling and the development of open-source datasets with standardized evaluation protocols to foster broader adoption and fair comparative studies. We also plan a systematic evaluation on downstream tasks, which will be crucial for demonstrating the practical value of the proposed speaker-aware simulation of multi-talker conversation timing beyond theoretical analyses.

\section*{Acknowledgment}
Project No. 2025-2.1.2-EKÖP-KDP-2025-00005 has been implemented with the support provided by the Ministry of Culture and Innovation of Hungary from the National Research, Development and Innovation Fund, financed under the EKÖP\_KDP-25-1-BME-21 funding scheme.

\fontsize{10pt}{0pt}\selectfont
\bibliographystyle{IEEEbib}
\bibliography{main}

\end{document}